\documentclass[aps, prl, twocolumn]{revtex4}
\usepackage{amsmath}
\usepackage{amssymb}
\usepackage{amsthm}
\usepackage{xspace}
\usepackage{graphicx}
\usepackage{paralist}
\usepackage[usenames,dvipsnames,table]{xcolor}

\newcommand{\ket}[1]{| #1 \rangle}

\newcommand{\inner}[2]{\langle #1, #2 \rangle}

\newcommand{\calA}{\mathcal{A}}
\newcommand{\calB}{\mathcal{B}}
\newcommand{\accept}{\mathrm{accept}}
\newcommand{\reject}{\mathrm{reject}}

\begin{document}

\title{\vspace{-1cm}Simple proof of the impossibility of\\ bit-commitment in generalised probabilistic theories\\ using cone programming}

\author{Jamie Sikora}
\affiliation{Perimeter Institute for Theoretical Physics, Waterloo, Ontario, Canada, N2L 2Y5}

\author{John Selby}
\affiliation{Perimeter Institute for Theoretical Physics, Waterloo, Ontario, Canada, N2L 2Y5}

\date{\today}

\begin{abstract}
Bit-commitment is a fundamental cryptographic task,
in which Alice commits a bit to Bob
such that she cannot later change the value of the bit, whilst,  simultaneously, the bit is hidden from Bob.
It is known that ideal bit-commitment is impossible within quantum theory.
In this work, we show that it is also impossible in generalised probabilistic theories (GPTs) (under a small set of assumptions) by presenting a quantitative trade-off between Alice's and Bob's cheating probabilities.
Our proof relies crucially on a formulation of cheating strategies as cone programs, a natural generalisation of semidefinite programs.
In fact, using the generality of this technique, we prove that this result holds for the more general task of integer-commitment.
\end{abstract}

\maketitle

The discovery of quantum theory in the early 20th century immediately and radically altered our understanding of the physical world. However, the consequences of this discovery have taken much longer to unravel. Indeed, it was only in the 1980s with the emerging field of quantum information theory, that the implications for information processing, computation, and cryptography started to become apparent.

Perhaps the most striking example is quantum key distribution.
The security of the distributed key is no longer contingent upon assumptions about the computational resources of an eavesdropper, but instead is based on the assumption that
quantum theory is a faithful description of nature.
This is a more solid foundation on which to base cryptographic security. However, it is plausible that quantum theory will one day be replaced by a more fundamental theory and so the security would have to be re-evaluated.
It is therefore highly desirable to base security not on the validity of quantum theory itself, but on basic physical principles that we may expect to hold even in a post-quantum world.
For the case of key distribution it has been demonstrated \cite{barrett2005no} that security can be based only on the principle of no-signalling. In other words, as long as nature does not allow for instantaneous communication then perfectly secure key distribution is possible.

In this paper we explore the cryptographic task of \emph{integer-commitment} which is
known to be impossible assuming only quantum theory~\cite{May97, LC97a}.
Our main result is that this also holds in \emph{generalised probabilistic theories}, abbreviated as \textup{GPT}s, satisfying
the No-Restriction Hypothesis and the Purification Postulate (which we define later).
Moreover, we give a quantitative trade-off between the extents to which Alice and Bob can cheat.
The case of bit-commitment--a special case of integer-commitment--was proven to be impossible in GPTs~\cite[Corollary 45]{chiribella2010probabilistic} satisfying a slightly different set of assumptions (they do not assume the No-Restriction Hypothesis but assume \emph{local discriminability} and \emph{the existence of perfectly discriminable states}).

The novelty of this work is the simplicity of the proof technique.
Our proof mainly relies on the use of \emph{cone programming}, also known as \emph{linear conic optimisation}, a tool which is perfectly suited to optimising quantities arising in the study of  GPTs.
However, to the best of our knowledge, there have only been a few papers which use cone programming in the setting of GPTs~\cite{fiorini2014generalized, JP17,bae2016structure,LPW17} and in the setting of quantum theory~\cite{GSU13, BCJRWY14, LP15, NST16, SW17}.
In this paper we provide an introduction to cone programming and GPTs to illustrate the close connections  between the two. Moreover, we demonstrate, along with the references above, that cone programming is a natural tool to be used in the study of GPTs.
We hope that the conjunction of the two fields will lead to new perspectives and results in physics and/or optimization theory.

\section{Introduction to GPTs}

The framework of GPTs \cite{hardy2001quantum,barrett2007information,barnum2011information,chiribella2010probabilistic} describes essentially any theory of nature admitting an operational description. It is based on the idea that ultimately any theory of nature is going to be understood in terms of experiments, and so should have a description in terms of these experiments.
Primitive experimental notions, such as classical control--that we can choose which experiment to perform--and tomography--that states can be characterised by experiments--lead to strong mathematical constraints on the theory.
See \cite{barrett2007information} for a pedagogical and well-motivated introduction to the framework. Here we just present the bare bones mathematical structure of GPTs.

Each system, $\calA$, in a GPT is described by a pair of finite-dimensional closed convex cones, one of states $K_\calA$, and one of measurement outcomes (called effects) $E_\calA$.
Moreover, the cone of effects comes with a unit element selected from the interior denoted $u_\calA \in \mathsf{int}(E_\calA)$. $u_\calA$ is interpreted as the unique deterministic effect which allows us to `discard' systems.
In the case of quantum theory, the state cone and the effect cone are both equal to the cone of positive semidefinite matrices and the unique deterministic effect $u$ is provided by the partial trace.
Also, we make the common assumption that $\mathsf{Dim}(K_\mathcal{A})=\mathsf{Dim}(E_\mathcal{A})$, ensuring that tomography is possible.

To discuss the relationships these cones share with one another, we need the concept of the dual cone.
For a cone $K$, its dual cone, denoted $K^*$, is the set of covectors that evaluate to the non-negative reals on every vector in $K$.
One can check that the dual cone is a closed convex cone.

We think of measuring the state $s \in K_{\calA}$ with an effect $\{ e_1, \ldots, e_n \} \subset E_{\calA}$ which yields the outcome ``$e_i$'' with probability $\mathbf{Pr}(e_i|s):=e_i[s]$
where $e_i[s]$ denotes evaluation of the effect $e_i$ on the state $s$.
For this to make sense, we impose certain requirements on the triple $(K_\calA, E_\calA, u_\calA)$:
\begin{enumerate}[i)]
\item The effect cone and the state cone are contained in the other's dual:
\begin{equation} \label{dualcones}
E_\mathcal{A} \subseteq K_\mathcal{A}^* \; \text{ and } \; K_\mathcal{A} \subseteq E_\mathcal{A}^*;
\end{equation}
\item The physical states are given by the convex set:
\begin{equation} \label{physicalstates}
\left\{ s \in K_\mathcal{A} : u_\mathcal{A}[s] = 1 \right\};
\end{equation}
\item The set of valid effects are given by the convex set:
\begin{equation}
\left\{ (e_1, \ldots, e_n) : e_1, \ldots, e_n \in E_{\calA}, \; \sum_{i=1}^n e_i = u \right\}.
\end{equation}
\end{enumerate}
The above three requirements ensure the outcome of measuring a state with an effect yields a proper probability distribution.
As such, any such triple $\mathcal{A}:=(K_\mathcal{A},E_\mathcal{A},u_\mathcal{A})$ defines a valid system.
In quantum theory, the physical states are unit trace positive semidefinite operators, i.e., density operators, and the effects are simply POVMs. Just as \emph{pure quantum states} are extremal in this set of density operators, we say that the pure states of a GPT are extremal in the set \eqref{physicalstates}.
Any physical state which is not pure is said to be \emph{mixed}.


A GPT is described by selecting a particular set of systems $\{\mathcal{A,B,C},...\}$ which is closed under forming composites. The composite of system $\mathcal A$ and system $\mathcal B$ is denoted $\mathcal A\otimes B$ where $\otimes$ is associative.
This composite of systems induces a composite of the associated state and effect spaces, that is, we define $K_\mathcal{A}\otimes K_\mathcal{B}:=K_{\mathcal{A\otimes B}}$, $E_\mathcal{A}\otimes E_\mathcal{B}:=E_{\mathcal{A\otimes B}}$ and $u_\mathcal{A}\otimes u_\mathcal{B}:=u_{\mathcal{A\otimes B}}$.
This composite is subject to the following constraints
\begin{enumerate}[i)]
\item Bilinearity (so that $v\otimes\cdot$ and $\cdot\otimes w$ are linear maps for any $v$ or $w$);
\item $e_\mathcal{A}\otimes f_\mathcal{B}[s_\mathcal{A}\otimes t_\mathcal{B}] = e_\mathcal{A}[s_\mathcal{A}] f_\mathcal{B}[t_\mathcal{B}]$,\\ for all $s_\mathcal{A}\in K_\mathcal{A}$, $t_\mathcal{B}\in K_\mathcal{B}$, $e_\mathcal{A}\in E_\mathcal{A}$, and $f_\mathcal{B}\in E_\mathcal{B}$.
\end{enumerate}
Note that the unit effects $u$ provide a unique way to construct marginal states~\footnote{This is often not assumed as part of the basic framework in which case it is known as the causality postulate \cite{chiribella2010probabilistic}.}, namely $s_\calA:=u_\calB [s_{\calA\calB}]$.
Indeed, in quantum theory, $\otimes$ is just the standard tensor product and we represent marginal quantum states using the partial trace.

For each pair of systems there is a set of transformations $T_{\mathcal{A\to B}}$.
We do not formally define these here, for brevity, but we refer the reader to \cite{barrett2007information} for details.
Simply note that they map states to states and can be applied locally to one part of a bipartite system.
A transformation is \emph{reversible} if it has an inverse that is also a transformation in the theory.

\medskip
\emph{Assumptions--}Aside from the general framework described above we make two assumptions about the GPT which we simply define here and further discuss later.

\smallskip
\noindent \textbf{No-Restriction Hypothesis \cite{chiribella2010probabilistic}:}
For all systems $\calA$, we have $E_\calA= K_\calA^*$ and $E_\calA^* = K_\calA$. 
\smallskip

In quantum theory we can see that this is indeed satisfied as the cone of positive semidefinite matrices is self-dual, thus  $E_\calB = E_\calB^* = K_\calB = K_\calB^*$.

\smallskip
\noindent \textbf{Purification Postulate \cite{chiribella2010probabilistic}:} For any system $\calB$ there exists a system $\calA$ such that any (potentially mixed) state $s_\calB$ has a \emph{purification} $\psi_{\calA\calB}$ which is a pure state of the bipartite system such that $u_\calA[\psi_{\calA\calB}]=s_\calB$.
(If $t_{\calA \calB}$ is not pure but satisfies $u_\calA[t_{\calA\calB}]=s_\calB$, then we call $t$ a \emph{dilation} of $s$.)
Moreover, purifications are \emph{essentially unique}: any two purifications $\psi_{\calA\calB}$ and $\phi_{\calA\calB}$ of the same state $s_{\calB}$ are related by a reversible transformation on the purifying system $\calA$.

\smallskip
In the Discussion section, we address how these assumptions, or ones similar, are required for our results. 
It is important however to note that these assumptions do not restrict the GPT to quantum theory. Indeed, quantum theory over the real numbers \cite{hardy2012limited} also satisfies these assumptions. Moreover, based on the work of  \cite{barnum2017ruling}, for example, it would be surprising if there were not many more theories fitting these assumptions as well.

\section{Introduction to Cone Programming}

Cone programming is a generalisation of linear programming and semidefinite programming which have each seen many uses in quantum theory.
Just as semidefinite programming is perfectly suited to optimising quantities arising in quantum theory, cone programming serves that purpose for GPTs.

Suppose we have two finite-dimensional real inner product spaces $\cal{V}$ and $\cal{W}$, two vectors $C \in \cal{V}$ and $B \in \cal{W}$, a linear function $\Phi:\cal{V} \to \cal{W}$, and a closed convex cone $K \subseteq \cal{V}$. Then the cone program associated to the data $(\Phi, B, C, K)$ is a convex program of the following form~\footnote{Any convex program can be written in the form~\eqref{primal} by choosing appropriate $(\Phi, B, C, K)$.}
\begin{equation} \label{primal}
\alpha = \sup \{ \inner{C}{X} : \Phi(X) = B, \; X \in K \}.
\end{equation}
Here $X$ is the variable we are optimising over, and we want it to make the inner product $\inner{C}{X}$ as large as possible, but we only care about such $X \in K$ that satisfy the constraint $\Phi(X) = B$.
Note that an optimal $X$ may not exist, even if $\alpha$ is finite, hence the need for ``sup'' instead of ``max''.

When $K$ is the cone of positive semidefinite matrices we recover semidefinite programming and when $K$ is the nonnegative orthant, we recover linear programming. Thus, cone programming generalises both of these well studied classes of optimisation problems as noted earlier.

Cone programming has a rich theory, we refer the interested reader to the book~\cite{BV} and the references therein.
We refer to the optimisation problem \eqref{primal} as the primal problem. To the primal problem we associate the dual problem, defined as
\begin{equation} \label{dual}
\beta = \inf \{ \inner{B}{Y} : \Phi^*(Y) = C + S, \; S \in K^* \}
\end{equation}
where $\Phi^*$ is the adjoint of $\Phi$.
Here, we optimise over both $Y$ and $S$ which can be thought of as Lagrange multipliers of the primal problem. $Y$ corresponds to the linear constraint $\Phi(X) = B$ and $S$ corresponds to the cone constraint $X \in K$.
We see that both the primal and dual problems are optimization problems over the data $(\Phi, B, C, K)$.

In this work we only use a very basic result in cone programming called \emph{strong duality}
\footnote{Note that this property is not related to the notion of \emph{strong self duality} which is a property of a GPT which states that the state cone and effect cone are isomorphic in a particular way.}, stated below.

\smallskip
\noindent \textbf{Strong duality:}
If $\alpha$ is finite, and there is an ${X \in \mathsf{int}(K)}$ such that $\Phi(X) = B$, then $\alpha = \beta$. Moreover, the dual attains an optimal solution, meaning there is a $Y$ and $S \in K^*$ such that $\Phi^*(Y) = C + S$ and $\inner{B}{Y} = \alpha = \beta$.
\smallskip

The significance of strong duality is that the dual optimal value equals the primal optimal value.
Therefore, when one only cares about the optimal value, we can replace one optimisation problem for the other. This often yields an entirely new perspective on the original problem.
This is indeed the approach we take in this paper, to give a novel perspective on cheating in cryptographic tasks such as integer-commitment.

\section{Integer-Commitment}

Bit-commitment is a fundamental cryptographic primitive in two-party cryptography.
For instance, it can be used to build other important  protocols such as coin-flipping~\cite{Blu81} and zero-knowledge proofs~\cite{GMR89}.
However, it is impossible to realise perfectly in a quantum world~\cite{May97, LC97a} (although imperfect protocols are now known~\cite{CK11}).

In this work we consider a more general protocol in which bits are replaced by integers. This protocol consists of two phases.
\begin{itemize}
\item Commit Phase: Alice chooses a uniformly random integer $j \in \{ 1, \ldots, n \}$ and creates the bipartite state $s^j \in K_{\cal{AB}}$ of the system $\calA \otimes \calB$. She commits $j$ to Bob by sending him the system $\calB$.
\item Reveal Phase: Alice reveals $j$ and sends the system $\calA$ to Bob. Bob applies the two-outcome effect
\begin{equation}
(e^j_{\accept}, e^j_{\reject})
\end{equation}
to check if the system $\calA \otimes \calB$ is in fact in state $s^j$.
\end{itemize}

If integer-commitment were possible, then, in between the two phases, Alice should be unable to alter the committed integer and Bob should be unable to learn anything about it.
Integer-commitment is also known to be impossible in quantum theory by the same arguments as for bit-commitment. Recently, quantitative lower bounds have been derived using the impossibility of a task known as die-rolling~\cite{AS10, Sik17}.

To study integer-commitment in the GPT framework, we do not assume any purity requirement on Alice's initial states, nor a projective/sharpness property on Bob's effects.
As a result, we do not assume the ``no honest abort'' assumption, that is, we allow for the case where Bob aborts the protocol even when Alice is honest.
We just assume that there exists an $\alpha > 1/2$ such that
\begin{equation} \label{honest}
e^j_{\mathrm{accept}}[s^j] \geq \alpha, \; \forall j.
\end{equation}
Indeed, $\alpha = 1/2$ is always achievable using a strategy for Bob that simply returns a random guess as his decision to accept Alice's integer or not. Therefore, it does not make sense to consider an effect that performs worse than this.
Note that in quantum theory, one often takes $\alpha = 1$, but we do not make this assumption.

A depiction of integer-commitment is presented below.

\begin{figure}[h]
  \includegraphics[width=\linewidth]{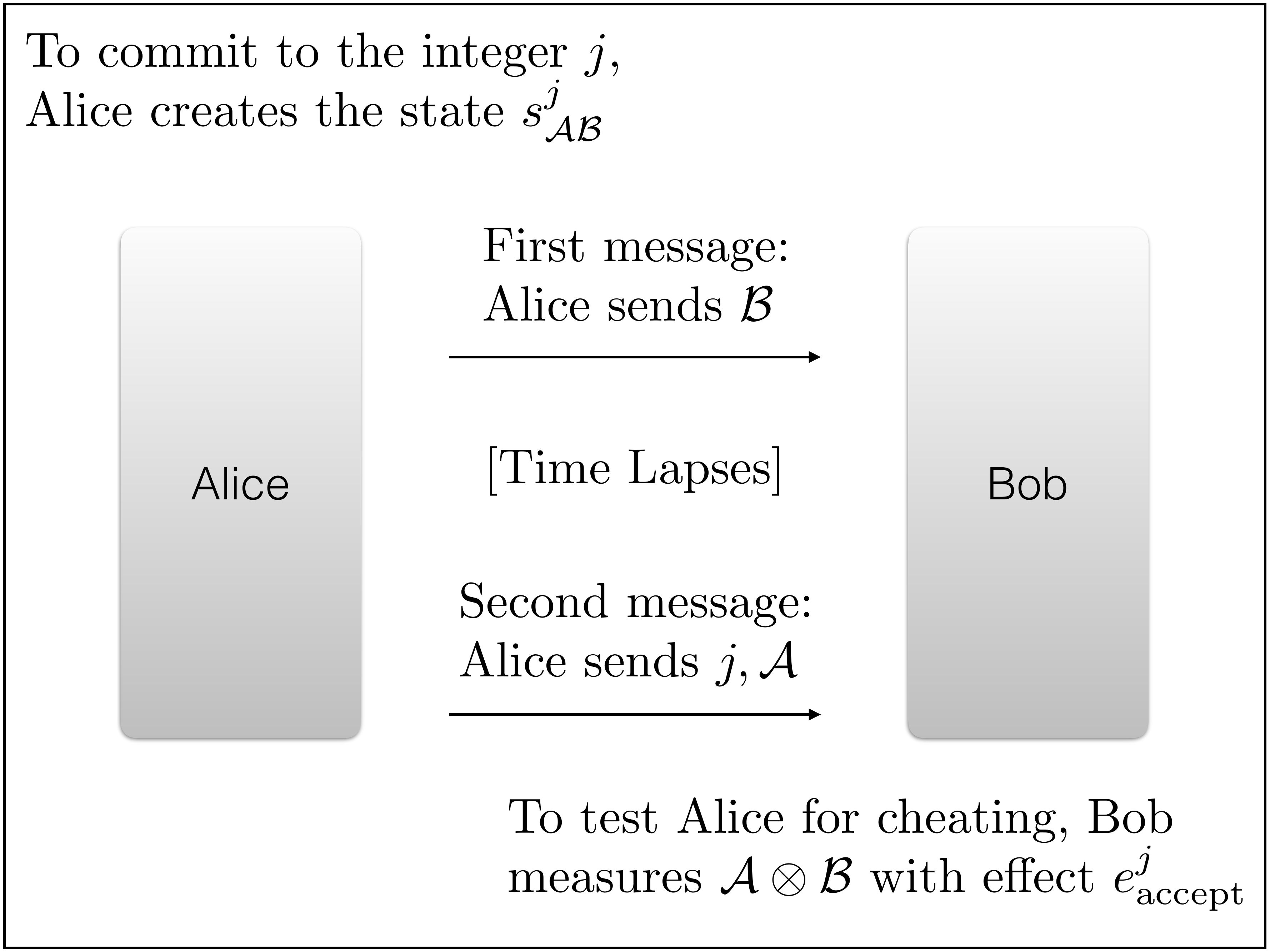}
  \caption{An integer-commitment protocol where Alice commits to an integer $j$ and Bob checks to see if Alice cheated.}
  \label{Fig_IC}
\end{figure}

We define the following symbols which capture the security notions of a given integer-commitment protocol:
\begin{itemize}
\item $P_B^*$: The maximum probability with which Bob can successfully learn the value of $j$ after the Commit Phase.
\item $P_A^*$: The maximum probability that Alice can successfully reveal an integer $j$ which is generated uniformly at random after the Commit Phase.
\end{itemize}

Bob's strategy is clear.
He must perform an effect $(f^1, \ldots, f^n)$ on his \emph{reduced state}
$\rho^j:=u_\calA[s^j]$ to learn $j$.
The maximum extent to which he can cheat is captured by the optimal objective value of the following cone program
\begin{equation} \label{BobPrimal1}
P_B^* = \sup \left\{
\frac{1}{n} \sum_{j=1}^n f^j [\rho^j] :
\sum_{j=1}^n f^j = u_\calB, \; f^j \in E_{\calB}, \; \forall j
\right\}.
\end{equation}

Since the cone program~\eqref{primal} is given in terms of vectors and not functionals, we use the Fr\'{e}chet-Riesz Representation Theorem which associates to each linear functional $\phi$ a unique vector $M$ such that $\phi(x) = \inner{M}{x}$ for all $x$ in the vector space.
Thus, Bob's cone program, in standard form, is given by
\begin{equation} \label{BobPrimal2}
\sup \left\{
\frac{1}{n} \sum_{j=1}^n \inner{M^j}{\rho^j} :
\sum_{j=1}^n M^j = U_\calB, \; \inner{M^j}{\cdot} \in E_{\calB}, \; \forall j
\right\}
\end{equation}
where $M^j$ is the unique vector associated to the effect $f^j$ and $U_\calB$ the unique vector associated to $u_\calB$.

At this point, we make use of the No-Restriction Hypothesis to assume $E_{\calB} = K_B^*$ (we later discuss why this, or some variant, is required).

It can be checked that the dual to~\eqref{BobPrimal2} is given as
\begin{equation} \label{BobDual}
\inf \left\{
u_\calB[x] : x - \frac{1}{n} \rho^j \in E_{\calB}^* = K_\calB, \; \forall j \right\}.
\end{equation}
We now want to check whether we can invoke strong duality.
Note that an interior point for the primal \eqref{BobPrimal2} is given by the effect which randomly outputs an outcome, i.e.,
\begin{equation}
M^j = \frac{1}{n} \, U_\calB, \;\forall j.
\end{equation}
Moreover, $P_B^*$ is obviously finite (since it is a probability) and thus we can apply strong duality to show that the optimal value of \eqref{BobDual} is in fact equal to the optimal value of \eqref{BobPrimal2}, namely $P_B^*$,  and also that the minimum is attained.
Thus, we can rewrite \eqref{BobDual} as
\begin{equation} \label{BobDual2}
P_B^* =
\min \left\{
u_\calB[x] : x - \frac{1}{n} \rho^j \in E_{\calB}^* = K_\calB, \; \forall j \right\}.
\end{equation}
We stress here that strong duality has given us a new perspective on a cheating Bob.
At first glance, there is no reason that the above optimisation problem should capture Bob's cheating probability.
We make use of this new perspective which, conveniently, uses the cone $K_{\cal{B}}$, the cone in which the marginals $\rho^j$ live.

We now describe a (not necessarily optimal) cheating strategy for Alice using this new insight.
For this, we use the Purification Postulate.
Let $x$ be an optimal solution to~\eqref{BobDual2} so that
\begin{equation} \label{BobOpt}
P_B^* = u_\calB[x] > 0.
\end{equation}
Notice that $x' = \frac{1}{u_\calB[x]} x \in K_\calB$ and $u_\calB[x']=\frac{u_\calB[x]}{u_\calB[x]}=1$ and so $x'$ is a valid normalised state in the GPT.
To see that $x'$ is indeed in the cone $K_\calB$ note that the constraint in~\eqref{BobDual2} implies that we can write
$x' = \frac{1}{u_\calB[x] \cdot n} (\rho^j + r^j)$ for some $r^j \in K_{\cal{B}}$.

Since $\rho^j$ is the marginal of $s^j$, we call $s^j$ a \emph{dilation} of $\rho^j$. In this sense,
let $t^j$ be a dilation of $r^j$. Then
\begin{equation} \label{chi}
\chi^j := \frac{1}{u_\calB[x] \cdot n} (s^j + t^j) \in K_{\calA \otimes \calB}
\end{equation}
is a dilation of $x'$ for each $j$. Thus, if we purify one of them to $\tilde{\chi}\in K_{\calA \otimes \calB \otimes \cal{C}}$ then it follows from the essential uniqueness condition in the Purification Postulate that $\tilde{\chi}$ can be transformed into any of the $\chi^j$ by first applying a reversible transformation on the composite system $\cal{A} \otimes \cal{C}$ and then discarding system $\cal{C}$.

We now have the following cheating strategy for Alice: Alice prepares the state $\tilde{\chi}$ and passes system $\calB$ to Bob keeping hold of the system $\cal{A} \otimes \cal{C}$.
Then to reveal $j$, she `steers' the state $\tilde{\chi}$ to $\chi^j$ using the procedure outlined above before sending $j$ and the system $\calA$ to Bob.

We can now compute the success probability of this strategy as
\begin{equation} \label{strat}
\frac{1}{n} \sum_{j=1}^n e^j_{\mathrm{accept}}[{\chi^j}]
\leq
P_A^*
\end{equation}
recalling that $P_A^*$ is Alice's maximum cheating probability.

By combining Eqs.~(\ref{honest}), (\ref{BobOpt}), (\ref{chi}), and (\ref{strat}), and the fact that $e^j_{\mathrm{accept}} [t^j] \geq 0$ as they live in dual cones, we arrive at our main theorem.

\smallskip
\textbf{Theorem.}
In any integer-commitment protocol, Alice and Bob's cheating probabilities satisfy
\begin{equation} \label{bound}
P_A^* \cdot P_B^*
\geq
\frac{\alpha}{n}
>
\frac{1}{2n}.
\end{equation}
This proves that integer-commitment is impossible since if Bob cannot cheat, i.e., $P_B \approx 1/n$, we have that Alice can cheat with probability at least $\approx 1/2$, making it insecure.
Note that if there is no honest abort probability, i.e., $\alpha = 1$, then Alice could cheat perfectly in this case.

\section{Discussion}

We now discuss and elaborate on the assumptions we used in our proof. Concerning the No-Restriction Hypothesis \cite{chiribella2010probabilistic}, something along these lines is necessary to ensure that Bob has sufficient measurements to suitably distinguish states. For example, if all of his measurements are constrained to be in a very small set,
distinguishing states would not be possible.
Though at first glance the No-Restriction Hypothesis seems not to be  very physically motivated, in various reconstructions of quantum theory it has been derived from natural physical principles \cite{chiribella2011informational,hardy2011reformulating,barnum2014higher,selbyReconstruction}.
Moreover, it is an important feature of both classical and quantum theory as well as more general theories based on Euclidean Jordan Algebras \cite{barnum2016composites,barnum2017ruling}.

Indeed, even in quantum theory if we restrict the effect cone such that the only allowable measurements on $\calB$ are trivial (multiples of the identity), and any POVM is allowable on $\calA \otimes \calB$, then the states $s^j = \ket{j}_{\calA} \ket{j}_{\calB}$ defines a protocol where neither Alice nor Bob can cheat.
Thus, the No-Restriction Hypothesis, or some variant, is necessary to prove that integer-commitment is impossible.

Concerning the Purification Postulate, this was first introduced in \cite{chiribella2010probabilistic} and aims to capture a principle of information conservation for a theory: that any lack of knowledge can ultimately be traced back to neglecting some system, whether it be an environment system or a system held by a space-like separated party.
It is known that entanglement is necessary for the impossibility of bit-commitment \cite{barnum2008nonclassicality} in any non-classical theory. Therefore, some principle is necessary beyond the No-Restriction Hypothesis to ensure the theory has entanglement. In our result we use the Purification Postulate to achieve this. However, it would be an interesting idea for future work to consider alternate principles such as those used in \cite{richens2017entanglement}.

\section{Conclusion}

We have shown that cone programming can be used to give a short and simple proof of the impossibility of integer-commitment, and therefore, bit-commitment, in any GPT satisfying the No-Restriction Hypothesis and the Purification Postulate.
We speculate that cone programming will have many future uses in the study of GPTs, just as semidefinite programming has been essential to the study of quantum information theory.
An immediate research direction which stems from this work is to determine which other cryptographic tasks are possible/impossible in the GPT framework.
Another would be to further study the GPT state discrimination problem, which is precisely the task of cheating Bob in this work.
In general, it will be exciting to see what other results of quantum theory can be extended to the GPT regime using cone programming.
By doing so, relaxing the assumption of the validity of quantum theory to the validity of basic physical principles that we could hope to be true regardless of the ultimate theory of nature.

\smallskip
\begin{acknowledgments}
\emph{Acknowledgements--}This research was supported in part by Perimeter Institute for Theoretical Physics.
Research at Perimeter Institute is supported by the Government of Canada through the Department of Innovation, Science and Economic Development Canada and by the Province of Ontario through the Ministry of Research, Innovation and Science.
\end{acknowledgments}

\bibliographystyle{plain}
\bibliography{bibliography}

\end{document}